\documentstyle[twocolumn,pre,aps]{revtex}
\begin{document}
\date{\today}
\draft
\title{Scaling Violations with Textures in Two-Dimensional Phase-Ordering}
\author{A. D. Rutenberg}
\address{Theoretical Physics Group, Department of Physics and Astronomy,
The University of Manchester, M13 9PL, UK}
\maketitle
\begin{abstract}
Scaling violations are found in the  phase-ordering two-dimensional
Heisenberg [$O(3)$] model, which has non-singular topological textures,
under dissipative non-conserved dynamics.
Three separate length-scales are found:
$L_T$ characterizes the scale of individual textures, $L_N$
characterizes the separation between textures, and
$L_C$ characterizes the distance between oppositely
charged textures.
\end{abstract}
\pacs{05.70.Ln, 64.60.Cn, 64.60.My}

Dynamical scaling has had great success in describing phase-ordering systems
with purely dissipative local dynamics \cite{Furukawa85}. The
scaling hypothesis is that the evolution of the system at late times
can be characterized by a single growing length-scale, $L(t)$, which scales
the two-point correlations of the system.
As yet, however, there is no general approach to determine whether a
particular system scales. Conserved spherical systems can be shown to
violate scaling \cite{Coniglio89}, while most other vector systems that
have been investigated seem to scale fairly well. Exceptions
are provided by one- and two- dimensional systems
with weakly interacting non-singular topological textures.  Textures
have both an internal scale characterizing the size of an individual
texture, and an external scale characterizing the texture separation.
If these scales evolve with different growth laws then scaling will be
violated.

The simplest system with topological textures is the one-dimensional (1D)
XY [$O(2)$] model, in which the textures are windings and anti-windings of the
XY phase along the system. These, and all other, textures are non-singular,
without the singular core structure of, e.g., vortices or domain walls.
In 1D, anomalous growth laws for both non-conserved and
conserved dynamics \cite{Newman90} indicate scaling violations
\cite{Rutenberg94a}
that have recently been observed and explained \cite{Rutenberg94b}.
In this paper, we explore two-dimensional (2D) systems
with textures, which we find also violate dynamical scaling.
Unlike the 1D systems, however, no extended scaling description is found to
apply to the two-point correlations.

2D Heisenberg [$O(3)$]  systems support topological textures  --- also
variously called skyrmions, instantons, or ``baby-skyrmions''.  An isolated
texture can be pictured as a stereographic projection of an order-parameter
sphere onto the plane of the system \cite{Rutenberg94a}.
The intrinsic scale
of the texture is then proportional to the radius of the projected sphere.
These 2D textures have recently been of particular interest: e.g.,
in cosmological skyrmionic strings \cite{Benson93},
in quantum antiferromagnets
\cite{Gooding94}, and also in particle physics \cite{Piette94}.
In all of these cases, 2D textures provide weakly-interacting,
localized, but non-singular, excitations.

Textures have an associated
topological charge which is quantized. After an arbitrary sign choice,
isolated textures have total charge $+1$, while isolated
anti-textures have total charge $-1$. Static solutions consisting
solely  of texture (or solely of anti-texture) configurations,
in a (hard-spin) non-linear sigma model,
were discussed by Belavin and Polyakov
\cite{Belavin75}. The solutions are
notable because the energy of the system is
independent of the overall scale and also of the
locations of the individual textures, each one of which
contributes unit charge and $8 \pi$ energy (using Eq. (\ref{EQ:HAM}), below).
These quasi-stable minimal
energy solutions demonstrate the weak interactions of 2D
textures.  On the other hand, systems with both textures and
anti-textures are never static \cite{Belavin75}.
However, the dynamics and evolution of generic textured systems
is relatively unexplored \cite{Piette94,Forster77}, particularly with
dissipative dynamics.  Indeed, the ``unwinding'',
or annihilation, of textures with anti-textures appears not to have been
addressed in two-dimensions. In this paper we consider mixed systems of
textures
and anti-textures developed by instantaneously quenching a disordered
state to $T=0$.  The phase-ordering dynamics are taken to be local,
purely dissipative dynamics.  For these systems, our numerical simulations find
a rich pattern of scaling violations.

The non-conserved ``model A'' dynamics are
\begin{equation}
\label{EQ:DYN}
	\partial_t \vec{\phi}({\bf x},t) = - \delta H/\delta \vec{\phi},
\end{equation}
where
\begin{equation}
\label{EQ:HAM}
	H = \int d^2 x \left[ ({\bf \nabla} \vec{\phi})^2 + V_0
		(\vec{\phi}^2-1)^2 \right].
\end{equation}
The lack of thermal noise in (\ref{EQ:DYN}) is appropriate for a quench
to $T=0$, where the equilibrium state has long-range order.
We simulate the dynamics numerically with both soft ($V_0 < \infty$) and
hard ($V_0=\infty$, or equivalently with a $|\vec{\phi}| =1$ constraint )
spins. We consider systems on square lattices of sizes
between $128 \times 128$ and $512 \times 512$, with periodic
boundary conditions  and independent randomly-oriented
unit-magnitude spins as initial conditions. We use at least $20$
independent runs for each system, and errors indicated in the figures
are extracted from the variations between runs. We use a simple Euler update
with a fixed time scale $\delta t = 0.01$ (except for $V_0=1/2$, where
we used $\delta t =0.1$), and consider times
up to $t = 10 000$. The late time regime covers nearly three decades of time,
starting with times $t \gtrsim 10$. This regime is unchanged with a smaller
timestep.  Similar results are obtained with different $V_0$,
although the early time behavior, and hence the asymptotic texture
density, changes if $V_0$ is small enough. In all cases,
however, the
asymptotic growth laws remain the same.

Previous numerical work has been carried out on these
2D $O(3)$ systems, using Eqs. (\ref{EQ:DYN}) and (\ref{EQ:HAM}),
with both hard-spins \cite{Bray90} and with soft-spins ($V_0=10$)
\cite{Toyoki93}, although topological quantities were not measured.
The hard-spin simulations, by Bray and Humayun,
found an energy density consistent with $\epsilon \sim t^{-2/3}$,
indicating a length-scale $L \sim t^{1/3}$, and no
dynamical scaling \cite{Bray90}.  The soft-spin simulations, by Toyoki,
found $L \sim t^{0.42 \pm 0.03}$ and dynamical scaling,
however only the energy and spin-spin correlations at
times $ t \lesssim 20$ (in our units) were investigated
\cite{Toyoki93}. In this
regime, we see strong early time transients.
Our results are consistent with this
previous work. However, we identify a distinct late time regime in which
scaling
is violated, and with growth law exponents that do not depend on $V_0$.

The topological texture density for these systems is given by
\begin{equation}
\label{EQ:TDENS}
	\rho({\bf x}) = [ \vec{\phi} \cdot ( \partial_x \vec{\phi} \times
					\partial_y \vec{\phi})]/4 \pi,
\end{equation}
which measures the local winding of the spin field around the unit sphere
in order-parameter space. We show two snapshots of the texture density for
part of
a single $512 \times 512$ system in figures \ref{FIG:SNAP1} and
\ref{FIG:SNAP2}.
For clarity, we only show texture density above the average magnitude in  each
figure. We identify the three evolving length-scales:
$L_T$ characterizes the scale of each texture,
$L_N$ characterizes the separation between textures independent of charge, and
$L_C$ characterizes the separation between oppositely charged textures.

Dimensionally the equation of motion (\ref{EQ:DYN}) determines a growing
length-scale $L \sim t^{1/2}$, where we have suppressed the dimensioned
kinetic coefficient. Growth laws different from this are only possible
through the introduction of other lengths: either the core scale $\xi \sim
V_0^{-1/2}$, or the initial correlation length $\xi_0$. We do not expect
any $\xi$ dependence in this system since there are
no singular defects.
Any $\xi_0$ dependence, on the other hand, indicates a scaling violation,
because then changing $\xi_0$ is not just equivalent to a shift in time
\cite{Rutenberg94a}. So, for systems without singular defects, any
growth law that differs from the dimensionally naive one indicates a scaling
violation. In addition, only positive powers of
$\xi_0$ can enter into any growth law, yielding smaller growth exponents
than $1/2$, because shorter initial correlations should only decrease
asymptotic length-scales.

We first show the energy-density and the average magnitude of the
texture density $\rho$ in Fig. \ref{FIG:LENGTHS}a.
The energy density $\epsilon$ follows from Eq.
(\ref{EQ:HAM}), and dimensionally scales as an inverse length-squared.
The texture density, from Eq. (\ref{EQ:TDENS}), has the same
scaling. For spins of unit magnitude, $\langle \rho \rangle$
is a conserved quantity \cite{lattice}.
However $\langle |\rho| \rangle$ is not,
and it decreases along with $\epsilon$ as textures unwind
with anti-textures.
At late times $\epsilon$ and $\langle | \rho | \rangle$
fall off as $t^{-0.65 \pm 0.02}$,
indicating a growing length scale $L_N \sim t^{0.33 \pm 0.01}$.  [We fit
times with $t>10$ and extract the approximate errors from the variance of the
exponents between different system sizes and different $V_0$.]
This length scale, $L_N$,
determines the overall texture density, and so characterizes
the separation of individual textures. There is
excellent quantitative agreement between the different system-sizes,
indicating that finite size effects only occur at later times,
and qualitative agreement
between the soft and hard-spin simulations at late
times. For soft-potentials, the spins are not saturated
at early times and this results in the initial
increase of the average magnitude of the
texture density seen for $V_0=1/2$, and the
shift in the asymptotic texture density at late times.

We can also consider the
portion of the energy contained in textures.  Since an isolated
hard-spin texture has an energy of $8 \pi$ using (\ref{EQ:HAM})
\cite{Belavin75}, we have plotted
$8 \pi \langle |\rho| \rangle$ in Fig. \ref{FIG:LENGTHS}a.
The agreement with $\epsilon$ at
late times indicates that the asymptotic energy-density is wholly contained
in the textures. The difference,
$\epsilon- \langle | \rho| \rangle$ (not shown),
decays very roughly like $t^{-1}$
--- and indicates a spin-wave like early time transient. This could explain why
the strong early-time correction to scaling observed
in $\epsilon$ is much smaller in the texture
density $\langle | \rho | \rangle$.

While we find no way to scale two-point correlations of either spins
or texture density, we can extract relevant length-scales from the evolution
of the correlations. However, because of the lack of scaling, this analysis
has significant systematic errors due to the evolving functional form of the
correlations.
In Fig. \ref{FIG:LENGTHS}b, we show the position of the first zero
of
\begin{equation}
\label{EQ:SIGNCORR}
	S(r,t) \equiv
	\langle \rho({\bf x}) \rho({\bf x}+{\bf r}) \rangle.
\end{equation}
This zero roughly
characterizes the texture-antitexture separation, $L_C$. The growth saturates
as $L_C$ approaches the system size. We find
$L_C \sim t^{0.4 \pm 0.1}$, fit over times $t>10$ and excluding late
times when finite size effects are apparent.

There is a heuristic  Energy-Scaling argument \cite{Rutenberg94a}
for $L_C \sim t^{1/2}$.  We first identify
the energy-density of the evolving system, $\epsilon \sim 1/L_N^2$.
Independently, we calculate the rate of energy-density dissipation,
$\partial_t {\epsilon}
= \langle \partial_t \vec{\phi} \; \delta H / \delta \vec{\phi}
\rangle  = - \langle (\partial_t \vec{\phi})^2 \rangle$, where we have used
(\ref{EQ:DYN}). Applying the dynamics to (\ref{EQ:HAM}), and neglecting
the potential term since it is subdominant energetically \cite{Rutenberg94a},
we find $\partial_t {\vec{\phi}} \simeq \nabla^2 \vec{\phi}$. Inside a texture,
$\nabla \vec{\phi} \sim 1/L_T$, while the Laplacian vanishes identically
for pure texture solutions \cite{Belavin75}. The natural length to enter
in the second derivative, then, is the texture-antitexture separation, $L_C$,
so we expect $\nabla^2 \vec{\phi} \sim 1/L_T L_C$. This will hold over the
area of the texture ($L_T^2$) for each texture (one per $L_N^2$), resulting in
$\partial_t {\epsilon} \sim 1/(L_T L_C)^2 L_T^2/L_N^2 \sim 1/ L_C^2 L_N^2$.
Comparing with $\epsilon \sim 1/L_N^2$, we find $L_C \sim t^{1/2}$.
This growth law is consistent with the data.

What of the third length-scale, $L_T$?  Since $L_T$ is the texture
scale, it characterizes the width of the peak centered at $r=0$ of the
two-point texture-density correlations. In Fig.
\ref{FIG:LENGTHS}c, we have plotted
the width at half-maximum of
\begin{equation}
\label{EQ:ABSTDENS}
A(r,t) \equiv \langle | \rho
( {\bf x})  \rho ({\bf x}+{\bf r}) | \rangle - \langle |\rho| \rangle^2.
\end{equation}
Fitting the points for $t>10$ we find
$L_T \sim t^{0.21 \pm 0.02}$, with similar results from $S(r,t)$.
We can check this growth law by
measuring $\langle \rho^2 \rangle$. Since there is one texture, of area
$L_T^2$ and with $\rho^2 \sim 1/ L_T^4$, in each area $L_N^2$, we expect
$\langle \rho^2 \rangle \sim 1/(L_N^2 L_T^2)$.  In Fig. \ref{FIG:LENGTHS}d, we
measure $\langle \rho^2 \rangle \sim t^{-1.02 \pm 0.03}$, by fitting $t>10$.
Using $L_N^2 \sim t^{0.65 \pm 0.02}$  from $\epsilon$, we have an independent
estimate of $L_T \sim t^{0.19 \pm 0.02}$. This is consistent, and we combine
these estimates to find $L_T \sim t^{0.20 \pm 0.02}$.

{}From the heuristic picture presented for $L_C$, small textures have larger
gradients and will annihilate earlier --- leading to an increasing $L_T$.
If we assume that the remaining textures have not evolved, then their internal
charge-density, $1/L_T^2$, is set by the initial fluctuations on the scale
$L_N$, of order $L_N/L_N^2$. This implies $L_T \sim L_N^{1/2}$. Using our
measurements for $L_T$ and $L_N$, it is then likely that $L_N \sim t^{1/3}$
and $L_T \sim t^{1/6}$ (where $\langle \rho^2 \rangle \sim t^{-1}$). These
exponents have been indicated by straight lines in Fig. \ref{FIG:LENGTHS}.

To summarize, we have found three characteristic length scales in the phase-
ordering of the 2D non-conserved Heisenberg model.
These length-scales exhibit different growth laws and so demonstrate the
violation of dynamical scaling in this system.
The growth laws are $L_C \sim t^{0.4 \pm 0.1}$ describing the separation of
textures and anti-textures,
$L_N \sim t^{0.33 \pm 0.01}$ characterizing the separation of textures, and
$L_T \sim t^{0.20 \pm 0.02}$ describing the scale of individual textures.
Soft and hard spin simulations give the same growth laws, despite individual
textures being destabilized by a finite $V_0$ \cite{Benson93,collapse}.
Similarly, the weak lattice instability \cite{lattice} does not seem
significant.  From our heuristic arguments, we
believe that $L_C \sim t^{1/2}$ exactly, and
our data is consistent with $L_N \sim t^{1/3} \xi_0^{1/3}$ and $L_T \sim
t^{1/6} \xi_0^{2/3}$. The factors of the initial correlation length $\xi_0$
make up the dimensions of length needed for the scaling violations.
A more convincing explanation for the growth laws
awaits a detailed understanding of the texture-antitexture unwinding
mechanism.

To what extent does this picture of scaling violations
carry over to other systems with non-singular topological textures? Certainly
in one-dimension a similar picture holds \cite{Rutenberg94b}.  We believe
that the observed scaling violations in 1D and 2D are due to the weak
interactions between textures.  This being the case, we would expect similar
scaling violations in conserved 2D $O(3)$ models, though these have not yet
been
explored to our knowledge.  However, because textures are strongly
unstable for $d>2$ \cite{Derrick64},
it would be surprising if similar scaling violations
were observed in higher dimensions.

\acknowledgments

I thank T. Blum, A. J. Bray, D. G. Clark, B. P. Lee, M. Monastyrsky, T. Newman,
and N. Turok for stimulating interactions, and also
the Isaac Newton Institute for hospitality during part of this work.

\begin{figure}
\caption{A snapshot of
one-quarter of a $512 \times 512$ system with $V_0=10$, at
$t=13.1$. Shown are the regions with texture density
above the average magnitude, $| \rho| \geq
\langle | \rho | \rangle$.
Black clusters have positive texture density, while outlined
grey clusters have negative texture density.
}
\label{FIG:SNAP1}
\end{figure}

\begin{figure}
\caption{A snapshot of the same system and region
as in the last figure, at $t=27.9$.
Again, we show $|\rho| \geq \langle |\rho | \rangle$.
}
\label{FIG:SNAP2}
\end{figure}

\begin{figure}
\caption{a) The energy density $\epsilon$ (top plots) and the texture
density $\langle |\rho| \rangle$ (bottom plots) vs. time after the quench.
The circles indicate indicate a system size
$L_\infty =512$ with $V_0 = 1/2$, the
squares indicate $L_\infty = 512$ with $V_0 = 10$, and
the asterisks indicate $L_\infty = 128$ with $V_0 = \infty$.
The straight line has slope $-2/3$. The dashed line is  $ 8 \pi
\langle |\rho| \rangle$ from the $L_\infty=512$ system with $V_0=10$.
b) The texture-antitexture separation $L_C$, characterized by
the position of the first zero of $S(r,t)$
vs. time after the quench. The symbols are the same,
with stars indicating $L_\infty=256$ with $V_0 =256$. The straight line
has slope $1/2$.
c) The texture size $L_T$, characterized by the
width, at half-height, of the first peak of $A(r,t)$ vs. time since the
quench. The straight line has slope $1/6$.
d) The average square texture density vs time since the quench. The
straight line has slope $-1$.
}
\label{FIG:LENGTHS}
\end{figure}


\begin{references}
\bibitem{lattice} We find that the lattice destabilizes a texture to
collapse. However only small textures
will collapse on the time-scales of our simulations,
and we see no qualitative changes of our results as the
average texture size grows.
\bibitem{collapse} Analytic and numerical treatment of isolated textures,
with $V_0 < \infty$, under
dissipative dynamics (\ref{EQ:DYN}) gives a scale $X(\tilde{t}) \sim
\tilde{t}^{1/4}$, where $\tilde{t}$ is the time remaining to collapse.
\end{references}
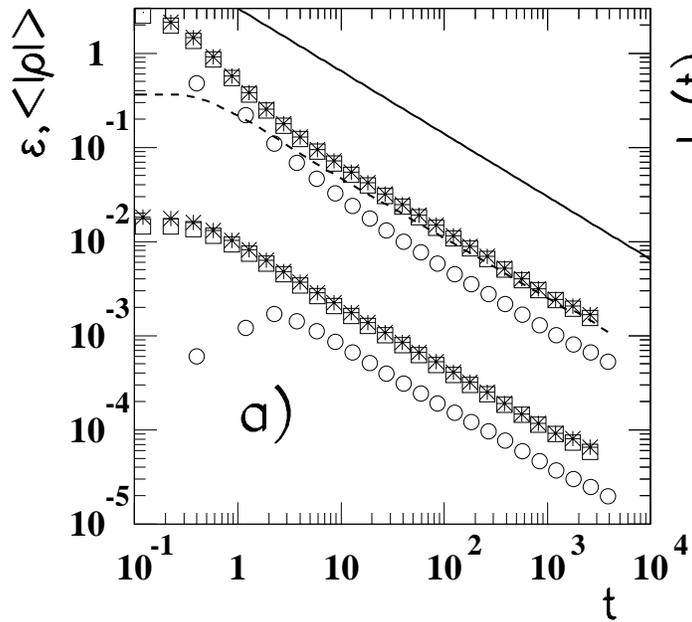
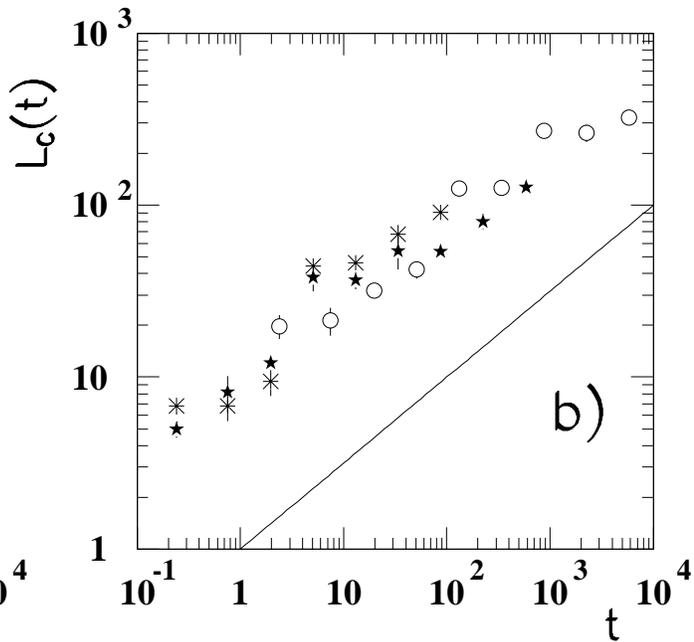
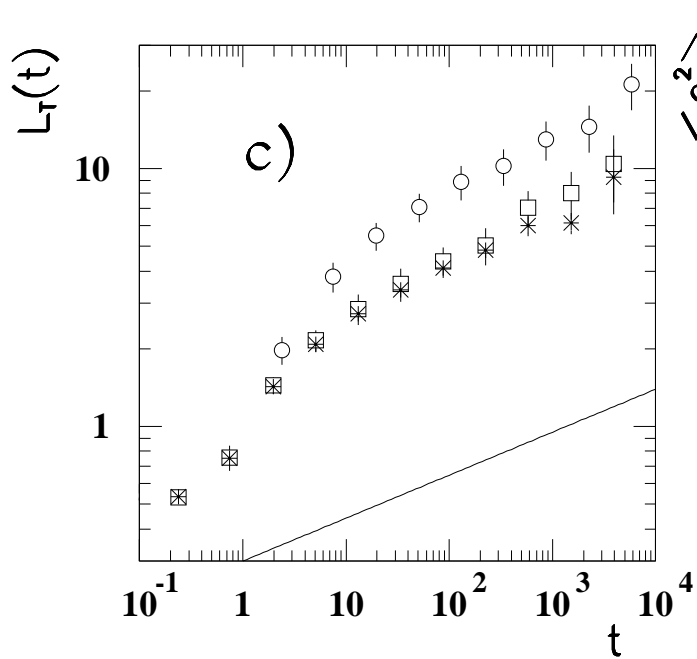
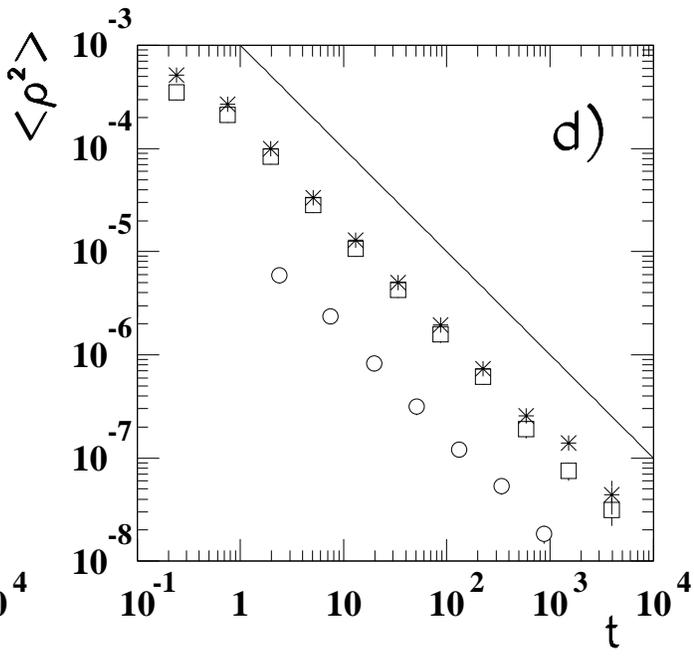
\end{document}